# Self Authentication of image through Daubechies Transform technique (SADT)

Madhumita Sengupta and J. K. Mandal
Department of Computer Science and Engineering,
University of Kalyani, Kalyani, Nadia-741235,
West Bengal, India
madhumita.sngpt@gmail.com
jkm.cse@gmail.com

*Abstract*—**In this paper a 4 x 4 Daubechies transform based authentication technique termed as SADT has been proposed to authenticate gray scale images. The cover image is transformed into the frequency domain using 4 x 4 mask in a row major order using Daubechies transform technique, resulting four frequency subbands AF, HF, VF and DF. One byte of every band in a mask is embedding with two/four bits of secret information. Experimental results are computed and compared with the existing authentication techniques like Li's method [5], SCDFT [6], Region-Based method [7] and other similar techniques based on Mean Square Error (MSE), Peak Signal to Noise Ratio (PSNR) and Image Fidelity (IF), which shows better performance in SADT.**

## I. INTRODUCTION

Digital medium stores massive amount of information in an incredibly less space. This is the main reliable way to protect information and preserve it for future use. The critical human psychologies impart brain to identify the object related information inside digital data. With a prior knowledge human eye is able to identify the actor and actresses from the running videos. Any audio can also be identified correctly by the name of the singer. But in general the paintings or photographs cannot hold the identification clue of the painter or the photographer.

Many techniques are available in today's digital world to protect ownership identity or to verify the authenticity of digital content both in spatial or frequency domain. The proposed work is a frequency domain based technique termed as SADT, where without preprocessing and use of any external information self authentication of digital image is achieved.

Various parametric tests are performed and results obtained are compared with existing techniques like Yuancheng Li's method [5], SCDFT [6], Region-Based method [7], SAWT [1], SADCT [2], IAHTSSDCT [3] and AWTDHDS [4] based on Mean Square Error (MSE), Peak Signal to Noise Ratio (PSNR) and Image Fidelity (IF) analysis [10] to show a consistent relationship with the quality perceived by the HVS (Human Visual System).

Section II deals with the overall scheme, the discrete Daubechies transforms technique[11] has been given in Section III. Technique for secret message computation is given in section IV. Embedding and authentication process is outlined in section V. Section VI analyzed the results of proposed work and comparisons to with existing techniques. Conclusions are drawn in section VII and that of references are cited at end.

## II. THE SCHEME

The proposed SADT technique is divided into two major tasks. Firstly at the source end the original image of dimension N x N passes through forward Daubechies transform FDT based on the mask of 4 x 4 window in a row major order to generate the corresponding frequency coefficients. Out of the four 'frequency coefficient' subbands the secret image feature in bits are extracted. Bits are then embedded into those frequency subbands, two/four bits per band of information based on the hash function. Schematic representation of encryption technique is shown in figure 1.

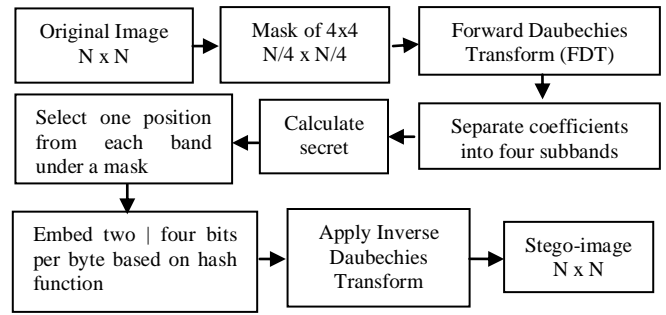

Figure 1. Schematic representation of encryption technique

At destination the stego-image of same dimension on receive passes through FDT to generate four frequency subbands. From every $3^{rd}$ coefficients of each band two/four secret bits are extracted to compare with image's own frequency feature computed at receiver end to authenticate. The schematic representation is shown in figure 2.

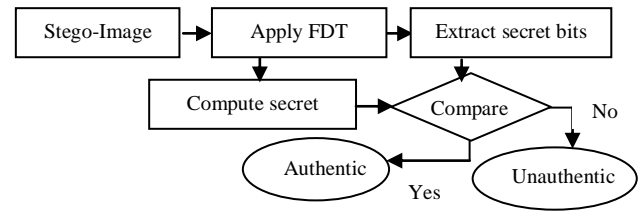

Figure 2. Schematic representation of encryption technique

## III. DISCRETE DAUBECHIES TRANSFORM TECHNIQUE

Every transformation technique comes with two sets of operations, forward transformation along with a pair of inverse transformation.

Daubechies wavelet has no explicit function expression. The scaling functions and wavelet functions are defined by the following two equations (1) and (2) respectively.

$$\emptyset(t) = \sum_{n=0}^{4} h[n].\emptyset(2t - n) \quad (1)$$

$$\psi(t) = \sum_{n=0}^{4} g[n].\emptyset(2t - n) \quad (2)$$

Where h[n] is a sequence of low-pass impulse response filter coefficients and g[n] is a sequence of high-pass impulse response filter coefficients, and the coefficients are:
h[0] = (1+SQRT (3))/ (4*SQRT (2)),
h[1] = (3+SQRT (3))/ (4*SQRT (2)),
h[2] = (3-SQRT (3))/ (4*SQRT (2)),
h[3] = (1-SQRT (3))/ (4*SQRT (2)),
g[0]= h[3], g[1] = -h[2], g[2] = h[1] and g[3]= -h[0].

On calculation of the forward Daubechies transform (FDT) the coefficients generated are grouped into four subbands, as shown in figure 3 and figure 4 respectively.

| Average / Lower Frequency (AF) | Horizontal / middle frequency (HF) |
|---|---|
| Vertical / middle frequency (VF) | Diagonal / Higher Frequency (DF) |

Figure 3.  Daubechies Coefficient in separate bands after FDT.

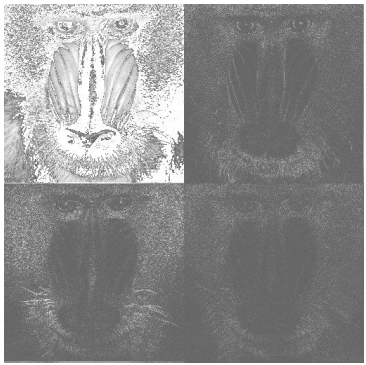

Figure 4.  Representation of subbands after FDT on Baboon image

Inverse Daubechies transform IDT is a similar operation as FDT, where frequency coefficients are transformed to spatial domain to generate stego-image on embedding.

## IV. SECRET MESSAGE COMPUTATION

Two sets of secret messages are generated in SADT. Set-1 having average of all the coefficients after FDT in a 4 x 4 mask by (3). Set-2 having average of only lower frequency band that is of 2 x 2 by (4). A hash function is used to embed the secret bits in $3^{rd}$ coefficients of every subband in each mask.

Hash computation formula H = (((column + row * 4) + 'total no of bits per band per mask) % 'maximum position allowed for embedding from LSB') is used.

$$Set_1 = \frac{1}{16} * \sum_{i=0}^{1}\sum_{j=0}^{1}(AF_{ij} + HF_{ij} + VF_{ij} + DF_{ij}) \quad (3)$$

$$Set_2 = \frac{1}{4} * \sum_{i=0}^{1}\sum_{j=0}^{1}(AF_{ij}) \quad (4)$$

## V. EMBEDDING AND AUTHENTICATION

Embedding and authentication are two separate operations supporting each other. Embedding is done for authentication and authentication is done at destination end to verify the originality.

In the process of embedding after FDT, $3^{rd}$ coefficient of every subband is used so that no two consecutive positions are tamper by secret bits. The positions selected for embedding in a single mask are $P_{10}$, $P_{12}$, $P_{30}$ and $P_{32}$ as shown in figure 5. For every mask set-1 computation generates 1 byte of secret information and set-2 generates another byte of information. That means on embedding set-1 secret information [8 bit per 16 bytes] the payload becomes 0.5 bpB while embedding set1 and set2 together, [16 bits per 16 bytes] payload becomes 1.0 bpB. Double information is embedded to verify the secret on retrieval and minimize the chance of attack.

| AF Band | $P_{00}$ | $P_{01}$ | $P_{02}$ | $P_{03}$ | HF Band |
|---|---|---|---|---|---|
|  | $P_{10}$ | $P_{11}$ | $P_{12}$ | $P_{13}$ |  |
| VF Band | $P_{20}$ | $P_{21}$ | $P_{22}$ | $P_{23}$ | DF Band |
|  | $P_{30}$ | $P_{31}$ | $P_{32}$ | $P_{33}$ |  |

Figure 5.  Single mask representation of four subbands

Embedding is done with simple computation of Anding/Oring operation.

## VI. RESULTS AND DISCUSSIONS

This section deals with the results of computation on embedding hidden data. Ten PGM [9] images have been taken and SADT is applied on each. All cover images are 512 x 512 in dimension as shown in figure 6.

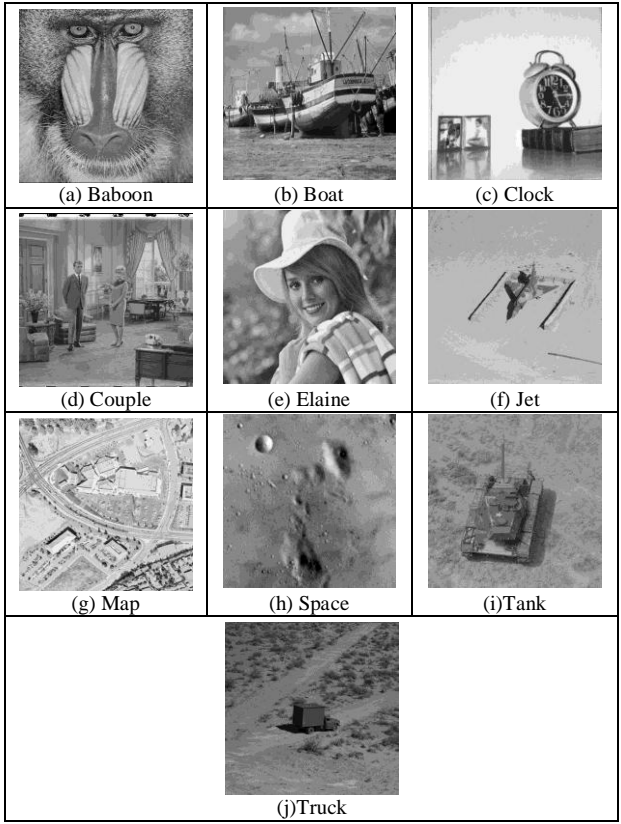

Figure 6.  Cover images of dimension 512 x 512

Calculation of average MSE, PSNR and IF for ten images with two bit secret per band per mask have been done to embed set-1 secret information and along with IDT. The average MSE for embedded image is 0.7006983 that of PSNR in dB is 49.692439 and IF is 0.999965 as shown in table I.

TABLE I.  DATA ON APPLYING SADT OVER 10 IMAGES WITH 0.5 BPB

| Cover Image 512 x 512 | MSE | PSNR | IF |
|---|---|---|---|
| (a) Baboon | 0.672668 | 49.852793 | 0.999964 |
| (b) Boat | 0.666504 | 49.892777 | 0.999965 |
| (c) Clock | 0.806648 | 49.063962 | 0.999979 |
| (d) Couple | 0.703968 | 49.655274 | 0.999958 |
| (e) Elaine | 0.693108 | 49.722797 | 0.999966 |
| (f) Jet | 0.649139 | 50.007424 | 0.999979 |
| (g) Map | 0.591949 | 50.407957 | 0.999983 |
| (h) Space | 0.770767 | 49.261571 | 0.999955 |
| (i) Tank | 0.678970 | 49.812296 | 0.999963 |
| (j) Truck | 0.773262 | 49.247537 | 0.999936 |
| *Average* | *0.7006983* | *49.692439* | *0.999965* |

Average of MSE, PSNR and IF for ten images with four bit secret per band per mask have been obtained on embedding set-1 and set-2 secret information. The average MSE after IDT is 1.6300945 and PSNR is 46.368404 and that of IF is 0.999916 as shown in table II. Figure 7.a shows the stego-image on embedding set-1 information with 0.5 bpB and figure 7.b shows stego-image on embedding set-1 with set-2 information with 1.0 bpB for robustness of the proposed SADT technique.

TABLE II.  DATA ON APPLYING SADT OVER 10 IMAGES WITH 1.0 BPB

| Cover Image 512 x 512 | MSE | PSNR | IF |
|---|---|---|---|
| (a) Baboon | 1.339035 | 46.862884 | 0.999928 |
| (b) Boat | 1.313438 | 46.946706 | 0.999931 |
| (c) Clock | 1.358120 | 46.801422 | 0.999964 |
| (d) Couple | 4.275150 | 41.821290 | 0.999744 |
| (e) Elaine | 1.329784 | 46.892991 | 0.999936 |
| (f) Jet | 1.427063 | 46.586372 | 0.999954 |
| (g) Map | 1.316666 | 46.936049 | 0.999961 |
| (h) Space | 1.299934 | 46.991589 | 0.999924 |
| (i) Tank | 1.308537 | 46.962945 | 0.999928 |
| (j) Truck | 1.333218 | 46.881793 | 0.999890 |
| *Average* | *1.6300945* | *46.368404* | *0.999916* |

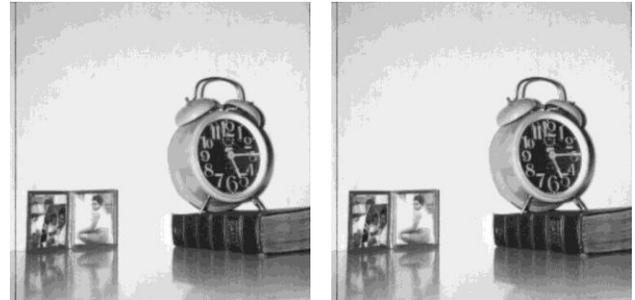

(a) Embedded with average of FDT mask (0.5 bpB payload)   (b) Embedded with average of FDT mask and LF band (1.0 bpB)

Figure 7.  Sample stego-image on embedding with payload 0.5 and 1.0 bpB.

TABLE III.  COMPARISON OF SADT WITH EXISTING TECHNIQUE

| Technique | Capacity (bytes) | Size of cover image | bpB (Bits per bytes) | PSNR in dB |
|---|---|---|---|---|
| Yuancheng Li's Method [5] | 1089 | 257 * 257 | 0.13 | 28.68 |
| SCDFT [6] | 3840 | 512 * 512 | 0.12 | 30.10 |
| SADCT [2] | 8192 | 512 * 512 | 0.08 | 56.63 |
| Region-Based [7] | 16384 | 512 * 512 | 0.5 | 40.79 |
| IAHTSSDCT [3] | 16384 | 512 * 512 | 0.5 | 47.48 |
| AWTDHDS [4] | 16384 | 512 * 512 | 0.5 | 44.87 |
| SAWT [1] | 131072 | 512 * 512 | 1.3 | 36.62 |
| SADT (set-1) | 16384 | 512 * 512 | 0.5 | 49.69 |
| SADT (set-1+set-2) | 32768 | 512 * 512 | 1.0 | 46.36 |

A comparative study has been made between Yuancheng Li's Method[5], SCDFT [6], SADCT[2], Region-Based method [7], IAHTSSDCT [3], AWTDHDS [4] and SAWT[1] with proposed SADT in terms of mean square error, peak signal to noise ratio and image fidelity. Comparison is done on average bases, table III and figure 8 display the comparison results in details. Comparison shows that proposed SADT generates optimised result in all respect.

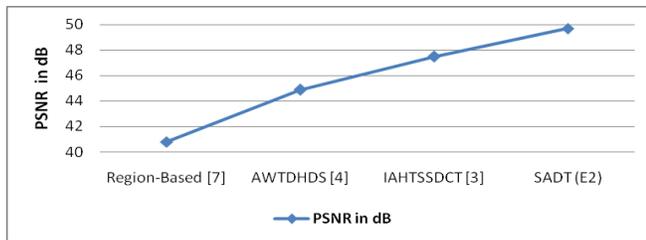

Figure 8.  Graphical representation of comparison for payload 0.5 bpB.

VII. CONCLUSION

SADT is a self embedding technique for authentication or ownership verification in a frequency domain based on Daubechies transform. This self authentication technique doesn't require any dictionary or code book or any preprocessing computation. SADT also works with optimal payload of 1.0 bpB without any noticeable changes by HVS.

ACKNOWLEDGMENT

REFERENCES

[1] Madhumita Sengupta, J.K. Mandal, "Self Authentication of Color image through Wavelet Transformation Technique (SAWT)", pp-151-154, ISBN 93-80813-01-5, ICCS 2010.

[2] Madhumita Sengupta, J. K. Mandal, "Self Authentication of Color Images through Discrete Cosine Transformation (SADCT)", IEEE catalog no : CFP1122P-CDR, ICRTIT, Anna University, Chennai, ISBN No-: 978-1-4577-0589-2, 3rd-5th June 2011.

[3] Madhumita Sengupta, J. K. Mandal "Image Authentication using Hough Transform generated Self Signature in DCT based Frequency Domain (IAHTSSDCT)",IEEE, ISED- 2011, Kochi, Kerala, pp- 324-328, DOI 10.1109/ISED.2011.43, 2011.

[4] Madhumita Sengupta, J. K. Mandal, "Authentication in Wavelet Transform Domain through Hough Domain Signature (AWTDHDS)" UGC-Sponsored National Symposium on Emerging Trends in Computer Science (ETCS 2012), ISBN number 978-81-921808-2-3, pp 61-65, 2012.

[5] Li Yuancheng, Xiaolei Wang, "A watermarking method combined with Radon transformand 2D-wavelet transform", IEEE, Proceedings of the 7th World Congress on Intelligent Control and Automation, June 25 - 27, Chongqing, China, 2008.

[6] T. T. Tsui, X. –P. Zhang, and D. Androutsos, Color Image Watermarking Usimg Multidimensional Fourier Transfomation, IEEE Trans. on Info. Forensics and Security, vol. 3, no. 1, pp. 16-28, 2008.

[7] A. Nikolaidis, I. Pitas, "Region-Based Image Watermarking", IEEE Transactions on Image Processing, Vol. 10, NO. 11, pp. 1721-1740, November 2001.

[8] J. K. Mandal, Madhumita Sengupta, "Authentication /Secret Message Transformation Through Wavelet Transform based Subband Image Coding (WTSIC)", IEEE, International Symposium on Electronic System Design 2010, pp 225-229, ISBN 978-0-7695-4294-2, Bhubaneswar, India, Print ISBN: 978-1-4244-8979-4, DOI 10.1109/ISED.2010.50, Dec, 20th -22nd, 2010.

[9] Allan G. Weber, The USC-SIPI Image Database: Version 5, Original release: October 1997, Signal and Image Processing Institute, University of Southern California, Department of Electrical Engineering. http://sipi.usc.edu/database/ (Last accessed on 25th May, 2011).

[10] M. Kutter , F. A. P. Petitcolas, A fair benchmark for image watermarking systems, Electronic Imaging '99. Security and Watermarking of Multimedia Contents, vol. 3657, Sans Jose, CA, USA, January 1999. The International Society for Optical Engineering, http://www.petitcolas.net/fabien/publications/ei99-benchmark.pdf. (Last accessed on 25th March, 2012).

[11]  Madhumita Sengupta, J. K. Mandal, "An Authentication Technique in Frequency Domain through Daubechies Transformation (ATFDD)", International Journal of Advanced Research in Computer Science (IJARCS), Volume 3 No. 4 (July-August 2012), www.ijarcs.info,  pp 236-242, ISSN No. 0976-5697.